\documentclass[letterpaper,aps,prd,preprint,showpacs,nofootinbib,
superscriptaddress]{revtex4-1}
\pdfoutput=1
\usepackage{epsf}
\usepackage{epsfig}
\usepackage{graphics}
\usepackage{graphicx}
\usepackage{amssymb}
\usepackage{color}
\usepackage{multirow}
\usepackage{amsmath}
\usepackage{float}

\newcommand{\nc}{\newcommand}
\nc{\postscript}[2]{\setlength{\epsfxsize}{#2\hsize}\centerline{\epsfbox{#1}}}

\nc{\beq}{\begin{equation}}   \nc{\eeq}{\end{equation}}
\nc{\bea}{\begin{eqnarray}}   \nc{\eea}{\end{eqnarray}}
\nc{\baa}{\begin{array}}      \nc{\eaa}{\end{array}}
\nc{\bit}{\begin{itemize}}    \nc{\eit}{\end{itemize}}
\nc{\ben}{\begin{enumerate}}  \nc{\een}{\end{enumerate}}
\nc{\bce}{\begin{center}}     \nc{\ece}{\end{center}}

\nc{\non}{\nonumber}

\begin{document}

\begin{flushright}

 \mbox{\normalsize \rm CUMQ/HEP 165}\\
        \end{flushright}
\vskip 20pt

\title {\bf  Saving the  fourth generation Higgs with radion mixing }

\vskip 20pt

\author{Mariana Frank\footnote{mfrank@alcor.concordia.ca}}
\affiliation{Department of Physics, Concordia University\\
7141 Sherbrooke St. West, Montreal, Quebec,\\ CANADA H4B 1R6
}
\author{
Beste Korutlu\footnote{beste.korutlu@gmail.com}}
\affiliation{Department of Physics, Concordia University\\
7141 Sherbrooke St. West, Montreal, Quebec,\\ CANADA H4B 1R6
}
\author{Manuel Toharia\footnote{mtoharia@physics.concordia.ca}}
\affiliation{Department of Physics, Concordia University\\
7141 Sherbrooke St. West, Montreal, Quebec,\\ CANADA H4B 1R6
}

\date{\today}

\begin{abstract}
We study Higgs-radion mixing in a warped extra dimensional model with
Standard Model fields in the bulk, and  we include a fourth
generation of chiral fermions.  
The main problem with the fourth generation is that, in the absence
of Higgs-radion mixing, it 
produces a large enhancement in the Higgs
production cross-section, now severely constrained by LHC
data. We analyze the production and decay rates of the two physical 
states emerging from the mixing and confront them with present LHC data.
We show that the current signals observed can be  
compatible
with the presence of one, or both, of these Higgs-radion mixed states
(the $\phi$ and the $h$),
although with a severely restricted parameter space. In particular,  
the radion interaction scale must be quite low, $\Lambda_\phi
\sim 1-1.3$ TeV. If $m_\phi \sim 125 $ GeV, the $h$ state must be
heavier ($m_h>320$). 
If $m_h \sim 125$ GeV,  the $\phi$ state must be quite light or 
close in mass ($m_\phi \sim 120$ GeV). 
We also present the modified decay branching ratios of the mixed
Higgs-radion states, including flavor violating decays into fourth
generation quarks and leptons.
The windows of allowed parameter space obtained are very sensitive to the
increased precision of upcoming LHC data.  During the
present year, a clear picture of this 
scenario will emerge, either confirming or further severely constraining this scenario.

\end{abstract}


\maketitle

\section{Introduction}

While the Standard Model (SM) is successful in explaining most, but not all, of the
present experimental data, it suffers from theoretical inconsistencies.
Outstanding amongst these are the two hierarchy problems:  the discrepancy
between the Planck and the electroweak scale, and the fermion mass hierarchy.
Thus it is generally expected that new physics around the TeV scale is needed to
stabilize the Higgs mass and solve the hierarchy problem. Originally, warped
extra dimensional models were introduced to explain the first 
discrepancy \cite {RS}. In the original scenario, two branes are introduced, one
with an energy scale set at the Planck scale, the other at the TeV scale,  with
the Standard Model (SM) fields localized  on the TeV brane, and with gravity
allowed to propagate in the bulk. The exponential warp factor arising from the
AdS geometry accounts for the seemingly unnatural difference between the Planck and the
electroweak scales.

Allowing  SM fermions and gauge fields to  
propagate in the bulk \cite{fermionsbulk} can also explain the fermion mass
hierarchy by fermion
localization \cite{a, bulkSM}.
One shortcoming of this minimal model is that the Kaluza-Klein (KK) excitations
of the bulk have masses compatible with the compactification scale, and 
 tight bounds from precision electroweak tests
\cite{EWPTmodel} and from flavor physics \cite{Weiler}
constrain them to be in the few TeV range, an obstacle in
producing and observing these resonances at the LHC.
 
The tests of the model would then likely come from observing other particles, in
particular, the radion, which  is a
scalar field associated with fluctuations in the size of the extra
dimension, and playing a role in its stabilization. In a simple model with a
bulk scalar which generates a vacuum expectation value (VEV), the
radion field emerges as a pseudo-Goldstone boson associated with
 the breaking of translation symmetry
\cite{Goldberger:1999uk}. The advantage is that the mass of the radion does not
depend on the compactification scale but only on the mechanism that
stabilizes the size of the extra dimension. As the radion couples more
or less through the trace of the energy-momentum tensor, its coupling
to particles is proportional to their mass, in a similar fashion to
the Higgs boson. In fact, as they share same quantum numbers, the
radion field can mix with the Higgs boson after electroweak symmetry
breaking, which involves another parameter, the coefficient 
of the curvature-scalar term \cite{GRW}.   Generically, the radion may be the lightest
new state in a Randall-Sundrum type setup, with a mass typically suppressed with
respect to KK fields by a volume factor of $\sim 40$ \cite{CGK}, which
then might put its mass between a few tens to hundreds of GeV, with
suppressed couplings which allow it to have escaped detection at LEP, and
consistent with precision electroweak data.

The interest in scalar particles beyond the SM has been fueled by the recent
ATLAS and CMS searches. Although clear evidence for a new state is at present 
inconclusive, there are exclusion regions reported by ATLAS  \cite{:2012si}: 
$112.9$ GeV $ \le m_{h^0} \le 115.5$ GeV, $131$ GeV $ \le m_{h^0} \le 238$ GeV and $251$ GeV $ \le m_{h^0} \le
466$ GeV at 95 \% confidence level (CL), while the mass region excluded by CMS 
\cite{Chatrchyan:2012tx}  is $127$ GeV $ \le m_{h^0} \le 600$ GeV at 95\% CL.  In the mean
time, both experiments observed an intriguing excess of events
for a possible Higgs  boson with mass close to $m_{h^0} = 126$ GeV for ATLAS and
$124$ GeV for CMS. The three most sensitive channels
in this mass range,  $h^0 \to \gamma \gamma, ~h^0 \to ZZ^* \to l^+l^-l^+l^-$,
and $h^0 \to WW^* \to l^+ \nu l^-  {\bar \nu}$  contribute to the excess with
significances of  3.1$\sigma$, 2.1$\sigma$ and 1.4$\sigma$, respectively.  If
this would be confirmed by further data analyses in 2012, a Higgs boson of mass
around 125 GeV may be indicative of physics beyond the SM due to the vacuum
instability \cite{Casas:1996aq}.
Furthermore the observed $3.1\sigma$ excess in the $\gamma \gamma$
 channel observed by ATLAS is higher than
the expected signals of the pure SM Higgs boson (with the same mass),
which again points to new physics.

As the window for new physics opens, one can naturally ask if the new state is a
Higgs, a radion, or a Higgs-radion mixed state. This possibility could have 
even more dramatic consequences for the scenario with an additional generation of fermions,
which is a natural extension of the warped space model as in
\cite{Burdman:2009ih}. Re-examination 
 of electroweak (EW) precision data showed that a fourth generation of fermions
is not ruled out experimentally. Additionally, an extra generation was shown  to
have theoretically
attractive features, and could help address some of the fundamental open
questions, such as baryon asymmetry of the universe, Higgs naturalness, fermion
mass hierarchy, and dark matter  (see \cite{holdom} and for a recent review,
\cite{Cetin:2011aa} and references therein).
However, at present four generations phenomenology may not be in such a good
shape. There are increasing indications from CMS and ATLAS  that the fourth
generation quarks would be very heavy \cite{Aad:2012bt}, though all the tests have
assumptions which may or may not hold. 
In any case, based basically on the enhancement of the Higgs production rate via $gg \to
h^0$ in four generations (by a factor of $4-9$), the ensuing enhancement
of the decay $h^0 \to gg$, and the suppression of $h^0 \to \gamma \gamma$
the Higgs boson is excluded in the mass range $120-600$
GeV at 95\%, and in the range $125-600$  GeV at 99\% C.L. \cite{CMS-12-008}.
In particular, it appears that if the bump in the signal at CMS and  LHC is the
Higgs boson, this would rule out the SM4 at 95\% confidence level for
$m_{h^0} \ge 123$ GeV, and at 99.6\% if $m_{h^0}=125$ GeV \cite{nir}. The
limits from the  Tevatron  \cite{Aaltonen:2010sv} also exclude a wide
range of Higgs boson masses.

In a recent study, we have shown that if the fourth generation is incorporated
into the framework of warped space models,  both the production and decay
patterns of the Higgs bosons can be altered significantly with respect
to the patterns expected in the standard model with four generations,
thus giving rise to distinguishing signals at the colliders
\cite{Frank:2011rw}. Radion phenomenology with and without Higgs-radion mixing
has been
discussed in several papers \cite{history,Dominici:2002jv,CHL}. It has been 
shown that a tree-level  misalignment between the flavor
structure of the Yukawa couplings of the radion and the fermion mass
matrix will appear when the fermion bulk parameters are not all
degenerate in both three \cite{Azatov:2008vm} and four generations
\cite{Frank:2011kz}. In this last reference it was also pointed out
that the radion is less sensitive to the presence of an extra
generation than the Higgs boson. 

As the mechanism responsible for the radion
flavor-changing neutral currents (FCNC's) is different from the one
for the Higgs  in these same models
\cite{Agashe:2009di,Azatov:2009na}, and the branching ratios for decays into gluons
and photons  for three and four generations also differs, we can expect the
phenomenology of the Higgs-radion mixed state to present an interesting
interplay of the two mechanisms responsible, and to yield different effects. In
particular, this mixing may help evade the apparent constraints on low
Higgs masses in the four generation scenario. 
Motivated by these expectations, we study the phenomenology of the Higgs-radion
mixed state, paying particular attention to the signals for $gg \to \phi \to
\gamma \gamma, ~gg \to \phi \to ZZ^*$ as well as $gg \to h \to \gamma \gamma,
~gg \to h \to ZZ^*$, where  $\phi$ and $h$ stand for the mixed Higgs-radion states. We use the presently-available ATLAS
\cite{:2012si}  and CMS data \cite{Chatrchyan:2012tx} for scalar
searches to identify regions in the parameter space where the data is
compatible with one or the other of these states.  Similar analysis with the mixing effects in three
generations has been most recently studied in
\cite{deSandes:2011zs}.

Our paper is organized as follows. In the next section (Sec.
\ref{sec:themodel}), we briefly review the warped model with fermions
in the bulk. We then discuss the production and decays of the
Higgs-radion mixed states (Sec. \ref{sec:prod-decays}) and discuss
regions of the parameter space where the events at the LHC  would be
compatible with observing a mixed state.  For the allowed parameter
space, we present the flavor-changing and flavor-violating decays of
the two Higgs-radion states in Sec. \ref{sec:FCNC}. We analyze cases
in which the fourth generation of leptons is $m_{\tau'}\sim 150$ GeV,
or light ($m_{\tau'}= 100$ GeV), as this mass has implication of
branching ratios. We summarize our findings and conclude in
Sec.\ref{sec:conclusion}.

\section{The model}
\label{sec:themodel}

The AdS metric including the scalar perturbation
$F$  is given in the RS
coordinate system by \cite{Charmousis:1999rg}
 \begin{equation}
ds^2 = e^{-2 (A + F)} \eta_{\mu\nu}dx^\mu dx^\nu - (1+2F)^2 dy^2
= \left(\frac{R}{z}\right)^2\left( e^{-2F} \eta_{\mu\nu} dx^\mu dx^\nu  -
(1+2F)^2 dz^2
\right),
\end{equation}
where $A(y) =k\,y$. 
The radion graviscalar is the scalar component of the
5D gravitational perturbations and  tracks fluctuations
of the size of the extra-dimension (i.e. its ``radius'').
The perturbed metric is no longer
conformally flat, and
in linear order in the fluctuation  $F$, the
metric perturbation is given by
\begin{equation}
\delta( ds^2 ) \approx -2 F \left( e^{-2A}
\eta_{\mu\nu} dx^\mu dx^\nu + 2~dy^2 \right) = -2 F \left( \frac{R}{z}
\right)^2 \left( \eta_{\mu\nu}dx^\mu dx^\nu + 2 dz^2\right),
\label{radionpert}
\end{equation}
where $F(z,x)$  is the 5D radion field. 
The radion acquires mass through a stabilization mechanism
with the addition of a bulk scalar
field with a vacuum expectation value (VEV), which, after taking into account the back-reaction
of the geometry due to the scalar field VEV profile~\cite{CGK}, leads to an effective
potential for the radion.  We assume that
this back-reaction is small,
and does not have a large effect on the 5D profile of the radion.

The relation between the
canonically normalized 4D radion field $\phi_0(x)$ and the metric
perturbation $F(z,x)$ is given by
\begin{equation}
F(z,x) = \frac{1}{\sqrt{6}} \frac{R^2}{R'} \left( \frac{z}{R}
\right)^2 \phi_0(x) =
\frac{\phi_0(x)}{\Lambda_\phi}\,\left(\frac{z}{R'}\right)^2,
\end{equation}
where $
\Lambda_\phi=\sqrt{6}M_{Pl}e^{-ky_{IR}}$ is the radion interaction
scale.\footnote{This relation of $\Lambda_\phi$ could be slightly  modified with the addition
  of gravity brane kinetic terms on the IR brane, and thus allow some
  flexibility on the precise definition of $\Lambda_\phi$ in terms of the other model parameters.}
We take the value of $k$ as $\sim M_5$, and $\Lambda_\phi\sim 
e^{-kL}M_5$ is expected to be $\sim$ TeV.
In the warped model with SM fields propagating in the bulk, radion
interactions with SM matter are slightly modified with respect to the
case of the SM lying on the TeV brane \cite{CHL}.
But still, 
they remain quite similar in form to the interactions of SM Higgs
except for an overall proportional constant, the inverse of the radion
interaction scale $\Lambda_\phi$.
The radion effective interaction Lagrangian yields the following
coupling to  gluons and photons: 
\begin{eqnarray}
{\cal L}_{g,A} &=& -\frac{\phi_0}{4\Lambda_\phi}\left[\left( \frac{1}{kL} +
\frac{\alpha_s}{2\pi}b_{QCD}\right) 
\sum_a G_{\mu\nu}^a G^{a\ \mu\nu} 
+\left( \frac{1}{kL} + \frac{\alpha}{2\pi}b_{EM} \right) 
F_{\mu\nu} F^{\mu\nu} \right]\ \ \ \ \ \ \ .
\label{eq2}
\end{eqnarray}
The radion couplings to $W$ and $Z$ bosons are
\begin{eqnarray}
{\cal L}_V &=& -\frac{2\phi_0}{\Lambda_\phi}\left[ \left( \mu_W^2W_\mu^+W^{-\
\mu}+\frac{1}{4kL}W_{\mu\nu}^+W^{-\ \mu\nu}  \right)
+\left( \frac{\mu_Z^2}{2}Z_\mu Z^{\mu}+\frac{1}{8kL}Z_{\mu\nu} Z^{\mu\nu} 
\right)\right],  \ \ \ \ \ 
\label{eq3}
\end{eqnarray}
where $V_{\mu\nu}=\partial_\mu V_\nu -\partial_\nu V_\mu$ for $V_\mu=W_\mu^\pm
,Z_\mu$ and
$\mu_i^2$ ($i=W,Z$)  include the contributions from the bulk wave functions of
$W,Z$, and 
are given as a function of the $W$ and $Z$ mass $m_{W,Z}$ 
$\displaystyle \mu_i^2=m_i^2\left [1-\frac{kL}{2}(\frac{m_i}{\tilde k})^2\right
]$.

The radion couplings to quarks (similar results hold for leptons) are
proportional to their masses:
\begin{eqnarray}
\label{Radioncoupling}
{\cal L}_f &=&-\frac{\phi_0(x)}{\Lambda_\phi} \left(q_L^{i} u_R^{j} +
\bar{q}_L^{i} \bar{u}_R^{j} \right)  {m^u_{ij}}
\left[{\cal I}({c_{q_i}}) + {\cal I}({c_{u_j}})\right] +( u \rightarrow d), 
\end{eqnarray}
where $\displaystyle \frac{c_{q_i}}{R}$,  $\displaystyle \frac{c_{u_i}}{R}$ and
$\displaystyle \frac{c_{d_i}}{R}$ are the 5D fermion
masses, and we choose to work in the basis where these are diagonal in 5D
flavor space. ${\cal I}(c)$ is obtained as
\begin{eqnarray}
\label{defI}
{\cal I}(c)= \left[\frac{(\frac{1}{2}-c)}{1-{(R/R')}^{1-2c}}+c\right]
\approx \Big\{ \begin{array}{c} c
  \, \,(\, c\, > \,1/2\,) \\ \frac{1}{2}\,\, (\, c \,<
  \,1/2\,) \end{array}.
\end{eqnarray}
This result was first obtained for the case of a brane Higgs and a single family
of fermions in \cite{CHL}, and it was later generalized to three
families and bulk Higgs in \cite{Azatov:2008vm}, where it was noted
that these couplings lead to flavor violation in the interactions
between the radion and fermions. 

Since the radion and the Higgs bosons have the same
quantum numbers, it is possible for them to mix via kinetic
factors:\footnote{We note that in the case of a bulk Higgs, there will
be Higgs-radion mixing  at the level of the bulk scalar potential, without the
need to introduce kinetic mixing. For simplicity, we will assume that the Higgs is
highly localized on the brane and consider only brane kinetic mixing.}
\bea
S_{\xi}=\xi \int d^4x \sqrt{-g_{vis}} R(g_{vis}){\hat H}^\dagger {\hat H}, 
\eea
with $R(g)$ the  Ricci scalar. The effective 4D Lagrangian up to quadratic order
will be
\begin{equation}
\mathcal{L}=-\frac{1}{2}\big(1+
6\gamma^2\xi\big)\phi_0\Box\phi_0-\frac{1}{2}\phi_0
m_{\phi_{0}}^2\phi_0-\frac{1}{2}h_{0}(\Box+m_{h_{0}}^2)h_{0}
-6\gamma\xi\phi_0\Box h_{0},
\end{equation}
where $m_{h_0}$ and $m_{\phi_0}$ are the Higgs and radion masses before mixing.
After rescaling to obtain states that diagonalize  the
kinetic and the mass terms (following \cite{Dominici:2002jv}) 
\bea
h_0&=&\left (\cos\theta -{6\xi\gamma/ Z}\sin\theta\right)h
+\left(\sin\theta+{6\xi\gamma\ Z}\cos\theta\right)\phi\equiv d h+c\phi
\label{hform}\\
\phi_0&=&-\cos\theta {\frac{\phi}{Z}}+\sin\theta {\frac{h}{Z}}\equiv
a\phi+bh\,.
\label{phiform}
\eea
with the mixing angle $\theta$ defined as
\beq
\tan 2\theta= 12 \gamma \xi Z
\frac{m_{h_0}^2}{m^2_{\phi_0}-m_{h_0}^2(Z^2-36\xi^2\gamma^2)}\,,
\label{theta}
\eeq
where
$\displaystyle
Z^2= 1+6\xi\gamma^2(1-6\xi)= \beta-36\xi^2\gamma^2\,$. The requirement
$Z^2>0$, which in turn gives $\beta>0$, is needed to maintain positive definite
kinetic terms for the physical fields $h$ and $\phi$. This requirement brings
theoretical limits on the $\xi$ parameter, which describes the mixing between
the Higgs and radion states, such that
\begin{equation}
\frac{1}{12}\Big(1-\sqrt{1+\frac{4}{\gamma^2}}\Big)\leq\xi\leq
\frac{1}{12}\Big(1+\sqrt{1+\frac{4}{\gamma^2}}\Big).
\end{equation}
The parameter $\xi$ is also subject to strong restrictions coming from
precision electroweak constraints (on $S$ and $T$ parameters),  LEP/LEP2 data,
and Tevatron bounds \cite{CGK,GTW}. In addition, there are theoretically excluded parameter
regions which do not satisfy requirements of $m_h-m_\phi$ degeneracy. The mass
squared values for the physical states are obtained as 
\begin{equation}
m_{\pm}^2=\frac{1}{2Z^2}(m_{\phi_{0}^2}+\beta
m_{h_{0}}^2\pm\sqrt{(m_{\phi_0^2}+\beta m_{h_0}^2)^2-4Z^2
m_{\phi_0^2}m_{h_0}^2}), 
\end{equation}
where the larger(smaller) of $m_h$ and $m_{\phi}$ will be identified as
$m_{+}(m_{-})$, and these must satisfy the inequality
\begin{equation}
\frac{m_{+}^2}{m_{-}^2}>1+\frac{2\beta}{Z^2}\Big(1-\frac{Z^2}{\beta}\Big)+\frac{
2\beta } { Z^2 } \Big(1-\frac{Z^2}{\beta}\Big)^{1/2}, 
\end{equation}
to keep the bare masses real. 

The presence of mixing will modify the couplings to fermions, gluons,
photons, $W's$ and $Z's$ of both the radion and the Higgs boson and
thus change the corresponding decay branching ratios as well as the 
production rates.

\section{Production and Decays of a mixed Higgs-radion State with four generations}
 \label{sec:prod-decays}

The main production mechanism of the Higgs particles at the hadron colliders is
through the gluon-gluon fusion channel, $\sigma(gg\rightarrow h_{SM})$, via
 triangular loops of heavy quarks. However, for heavier 
Higgs bosons, the weak vector boson fusion channel, $\sigma(qq\rightarrow
qqh_{SM})$, becomes competitive with the gluon-gluon fusion mode. Therefore, as a
good approximation one can write the ratio of the production cross
section of the $h$ physical mode to the production cross section of SM Higgs as follows
\begin{eqnarray}
 \frac{\sigma(gg\rightarrow h)+\sigma(qq\rightarrow qqh)}{\sigma(gg\rightarrow
h_{SM})+\sigma(qq\rightarrow
qqh_{SM})} =
 \bigg(\frac{\sigma(gg\rightarrow h)}{\sigma(gg\rightarrow
h_{SM})}+\frac{\sigma(qq\rightarrow qqh)}{\sigma(gg\rightarrow
h_{SM})}\bigg)\bigg(\frac{1}{1+\frac{\sigma(qq\rightarrow
qqh_{SM})}{\sigma(gg\rightarrow
h_{SM})}}\bigg), \nonumber \\
\label{HiggsProduction}
\end{eqnarray}

The ratio of the Higgs production cross section via the weak vector boson
fusion channel to the production cross section of the SM Higgs is closely
correlated with the partial widths such that 
\begin{equation}
 \frac{\sigma(qq\rightarrow qqh)}{\sigma(qq\rightarrow
qqh_{SM})}=\frac{\Gamma(h\rightarrow WW)}{\Gamma(h_{SM}\rightarrow WW)}~,
\end{equation}
which in warped extra dimensional scenarios with Higgs-radion mixing and fields
in the bulk simply becomes
\begin{equation}
\frac{\Gamma(h\rightarrow WW)}{\Gamma(h_{SM}\rightarrow
WW)}=\bigg(d+b\gamma(1-3\ln(\frac{\sqrt{6}M_{Pl}}{\Lambda_{\phi}})\frac{
M_W^2}{\Lambda_{\phi}^2})\bigg)^2.
\end{equation}
Substituting this result in Eq. \ref{HiggsProduction} we obtain, for the r.h.s., 
\begin{equation}
\label{hProduction}
\bigg[\frac{\sigma(gg\rightarrow h)}{\sigma(gg\rightarrow
h_{SM})}+\Big(d+b\gamma(1-3\ln(\frac{\sqrt{6}M_{Pl}}{\Lambda_{\phi}})\frac{M_W^2
} { \Lambda_ { \phi } ^2 }
)\Big)^2\frac{\sigma(qq\rightarrow qqh_{SM})}{\sigma(gg\rightarrow
h_{SM})}\bigg]\bigg(\frac{1}{1+\frac{\sigma(qq\rightarrow
qqh_{SM})}{\sigma(gg\rightarrow
h_{SM})}}\bigg),
\end{equation}
where the first term in the brackets is simply the ratio of couplings
to gluons $c_g^2/c_{g_{SM}}^2$. 

Similarly we can calculate the same ratio for the field $\phi$,
\begin{equation}\label{phiProduction}
\bigg[\frac{\sigma(gg\rightarrow \phi)}{\sigma(gg\rightarrow
h_{SM})}+\Big(c+a\gamma(1-3\ln(\frac{\sqrt{6}M_{Pl}}{\Lambda_{\phi}})\frac{M_W^2
} { \Lambda_ { \phi } ^2 }
)\Big)^2\frac{\sigma(qq\rightarrow qqh_{SM})}{\sigma(gg\rightarrow
h_{SM})}\bigg]\bigg(\frac{1}{1+\frac{\sigma(qq\rightarrow
qqh_{SM})}{\sigma(gg\rightarrow
h_{SM})}}\bigg).
\end{equation}
The production mechanism of an unmixed Higgs boson through the gluon-gluon fusion channel
increases about nine times with an additional fourth family of fermions, because in
addition to the top quark there are also heavy $t'$ and $b'$ quarks propagating
in the loop. Recently, the two-loop EW corrections, $\delta_{EW}^4$, to the
Higgs boson production via gluon-gluon fusion has been computed with respect to
the leading order cross section in 
\cite{Passarino:2011kv,Actis:2008ts,Actis:2008ug,Denner:2011vt}
\begin{equation}
 \sigma_{SM4}(gg\rightarrow h_0)=\sigma_{SM4}^{LO}(gg\rightarrow
h_0)(1+\delta_{EW}^4).
\end{equation}
This enters  as a correction of Higgs field prior to mixing. 

Also, in order to take into account the effects of KK fields in the loop, we assume an
additional correction to the $h_{0}$ couplings squared to massless gauge bosons of
$\pm 20\%$ for gluons and $\pm 10\%$ for photons. These corrections
can give enhancements or suppresions in the rates depending on the phases present at the level of the 5D
Yukawa couplings even if the effective quark masses and
mixings have the correct values for all parameter space points \cite{Azatov:2010pf}. In the figures, the effect will
be illustrated with bands in parameter space representing this
``theoretical uncertainty''.

With these considerations the $gg$ and
$\gamma\gamma$ couplings of the physical Higgs and radion fields become
\begin{eqnarray}
c_{g}^{h,\phi}(max)&=&-\frac{\alpha_{s}}{4\pi\upsilon}\Big[g^{h,\phi}_{g}(max)
\sum_i
F_{1/2}(\tau_i)-2\bigg(b'_{3}+\frac{2\pi}{\alpha_{s}\ln(\frac{\sqrt{6}M_
{ Pl } } {
\Lambda_ { \phi } } ) } \bigg)g^{h,\phi}\Big], \nonumber\\
c_{g}^{h,\phi}(min)&=&-\frac{\alpha_{s}}{4\pi\upsilon}\Big[g^{h,\phi}_{g}(min)
\sum_i
F_{1/2}(\tau_i)-2\bigg(b'_{3}+\frac{2\pi}{\alpha_{s}\ln(\frac{\sqrt{6}M_
{Pl}}{
\Lambda_ { \phi } } ) } \bigg)g^{h,\phi}\Big], \nonumber\\
c_{\gamma}^{h,\phi}(max)&=&-\frac{\alpha_{s}}{2\pi\upsilon}\Big[g^{h,\phi}_{
\gamma}(max) \sum_i
e_{i}^2N_{c}^{i}F_{i}(\tau_i)-\bigg(b'_{2}+b'_{Y}+\frac{2\pi}{\alpha_{s}
\ln(\frac { \sqrt { 6 } M_ { Pl } } {
\Lambda_ { \phi } } ) } \bigg)g^{h,\phi}\Big], \nonumber\\
c_{\gamma}^{h,\phi}(min)&=&-\frac{\alpha_{s}}{2\pi\upsilon}\Big[g^{h,\phi}_{
\gamma}(min) \sum_i
e_{i}^2N_{c}^{i}F_{i}(\tau_i)-\bigg(b'_{2}+b'_{Y}+\frac{2\pi}{
\alpha_{s}
\ln(\frac{\sqrt{6}M_{Pl}}{
\Lambda_ { \phi } } ) } \bigg)g^{h,\phi}\Big],\nonumber \\
\end{eqnarray}
where $b'_{3}, b'_{2}$ and $b'_{Y}$ are the coefficients of the beta
functions of the $SU(3), SU(2)$ and $U(1)_Y$ groups respectively,
in the presence of 4 generations of quarks and leptons. The coefficients are $b'_{3}=18/3$ and $b'_{2}+b'_{Y}=-65/9$.
Also, we have defined
\begin{eqnarray}
g^{\phi}_{g}(max)&=&a\gamma
+c\sqrt{(1+\delta_{EW}^4)(1.20)},\qquad
g^{\phi}_{\gamma}(max)=a\gamma +c\sqrt{(1.10)},\nonumber\\
g^{h}_{g}(max)&=&b\gamma
+d\sqrt{(1+\delta_{EW}^4)(1.20)},\qquad g^{h}_{\gamma}(max)=b\gamma
+d\sqrt{(1.10)},\nonumber\\
g^{\phi}_{g}(min)&=&a\gamma +c\sqrt{(1+\delta_{EW}^4)(0.80)},\qquad
g^{\phi}_{\gamma}(min)=a\gamma +c\sqrt{(0.90)},\nonumber\\
g^{h}_{g}(min)&=&b\gamma+d\sqrt{(1+\delta_{EW}^4)(0.80)},\qquad
g^{h}_{\gamma}(min)=b\gamma+d\sqrt{(0.90)},\nonumber\\
g^{\phi}&=&a\gamma ,\nonumber\\
g^{h}&=&b\gamma.
\end{eqnarray}
We mostly focus on the decays of Higgs-radion mixed
states to $\gamma\gamma$  and $ZZ^*$ for the low mass region,
and to $ZZ$ channel for larger masses. The ratio of discovery significances for
both the $h$ and the $\phi$ with respect to the SM Higgs can be defined as
\begin{equation}
 R_{h}(XX)=\Big[\frac{\sigma(gg\rightarrow h)+\sigma(qq\rightarrow
qqh)}{\sigma(gg\rightarrow h_{SM})+\sigma(qq\rightarrow
qqh_{SM})}\Big]\frac{BR(h\rightarrow XX)}{BR(h_{SM}\rightarrow
XX)}\ w_{corr}(h),
\end{equation}
and
\begin{equation}
 R_{\phi}(XX)=\Big[\frac{\sigma(gg\rightarrow \phi)+\sigma(qq\rightarrow
qq\phi)}{\sigma(gg\rightarrow h_{SM})+\sigma(qq\rightarrow
qqh_{SM})}\Big]\frac{BR(\phi\rightarrow XX)}{BR(h_{SM}\rightarrow
XX)}\ w_{corr}(\phi),
\end{equation}
where the terms in square brackets are defined in
Eqs.~(\ref{hProduction}) and (\ref{phiProduction}) and where
\bea
w_{corr}(s)= \left\{ \begin{array}{cc}\displaystyle \sqrt{\frac{\max(\Gamma_{tot}(h_{SM}),\Delta M_{4l})}{\max(\Gamma_{tot}(s),
\Delta M_{4l})}} \ \ \ \ & {\rm for}\ \ \
\Gamma_{tot}(s)>\Gamma_{tot}(h_{SM}) \\
\hspace{1.5cm} 1  \ \ \ \ & {\rm for}\ \ \ \Gamma_{tot}(s)<\Gamma_{tot}(h_{SM}).
 \end{array} \right.
\eea
The term $w_{corr}$ represents a crude approximation of the effects of a
large width of either $s=h$ or $s=\phi$. Indeed if the physical
state $h$ (or $\phi$) has a much larger width than the SM Higgs, and if
this width is larger than the experimental resolution of the
detector, then an LHC search looking for the SM Higgs would somewhat
underestimate the integrated signal as this one would be distributed
in a much wider resonance.

Finally, the experimental resolution in the 4-lepton channel is estimated to be \cite{GRW}
\begin{equation}
\frac{\Delta
M_{4l}}{M_{4l}}=\frac{0.1}{\sqrt{M_{4l}({\rm GeV})}}+0.005,
\end{equation}
We use all this information to explore the parameter space for
$m_\phi$ and $m_h$ consistent with  the LHC data, which indicates an excess in the
mass region $120-128$ GeV. 

We review the data so far. ATLAS data indicates an enhanced signal in $\gamma
\gamma$ and $ZZ^* \to 4\ell$ near $125$ GeV  \cite{:2012si} with observed excesses: $R(\gamma \gamma)= 2^{+0.8}_{-0.8},~ R(4 \ell)=0.5^{+1.5}_{-0.5}$. 

CMS data \cite{Chatrchyan:2012tx} indicates an excess at $124$ GeV:  $R(\gamma
\gamma)= 1.7^{+0.8}_{-0.7},~ R(4 \ell)=0.5^{+1.1}_{-0.5}$, and possibly an
additional enhancement either at $120$ GeV in $ZZ^*$ only: $R(4
\ell)=2^{+1.5}_{-1}$, or at  $137$ GeV in $\gamma \gamma$, $R(\gamma \gamma)=
1.5^{+0.8}_{-0.8}$ but not in $ZZ^*$, $R(4 \ell)<0.2$. The errors bars on the data
are still large, but they can be used to restrict the parameter for the four generation
Higgs-radion mixed states. In order for these states to fit the data, we should
either have one of the states at $124-126$ GeV, and another one hidden ({\it
i.e.} below the LHC signal), or one state at $124$ GeV and the other either at $120$ or $137$
GeV, both which should respect the CMS signal characteristics. 

Based on the experimental constraints, we investigate the production and decay of the two
scalar particles in our scenario, $m_\phi$ and $m_h$, and divide the
parameter space as follows. In the first scenario, we attempt to fit
$h$ as the scalar particle observed at LHC at an invariant mass of
$\sim125$ GeV, while requiring $\phi$ to
be consistent with constraints of the rest of the spectrum from LEP,
Tevatron and/or LHC; while in the second scenario, we attempt the same
thing for $\phi$, while $h$ must be consistent with the previous collider
data. 
\begin{itemize}
\item Scenario 1a: $m_h=124$ GeV, $m_\phi$ light ($<300$ GeV); in particular, paying specific
attention to $m_\phi=120$ GeV, $m_\phi=137$ GeV, as these seem possible parameter space points for the CMS data. 

\item Scenario 1b: $m_h=125$ GeV, $m_\phi$ heavy ($>300$ GeV).

\item Scenario 2a: $m_\phi=125$ GeV, $m_h$ light ($<300$ GeV); in particular, paying specific
attention to the point $m_h=120$ GeV.

\item Scenario 2b: $m_\phi=125$ GeV, $m_h$ heavy ($>300$ GeV).
 
\end{itemize}

 We illustrate some regions of parameter space with different
 masses of $h$ and $\phi$ in the following figures. The results
 will depend on the mass of the fourth family charged lepton
 ($\tau'$) and so we divide our considerations into two parts. We first
 assume that $m_{\tau'} \ge 150$ GeV, thus preventing
 flavor-changing decays into $\tau' \tau $, which are potentially
 large in this model \cite{Frank:2011rw, Frank:2011kz}. However, if
 the $\tau'$ is light, this might modify substantially the branching ratios, potentially
 yielding significantly different signals We comment on this case in
 this section, and investigate it in more detail in the next section.

\begin{figure}[t]
\center$
	\begin{array}{cc} 
\hspace*{-0.4cm}
	\includegraphics[width=2.6in,height=2.1in]{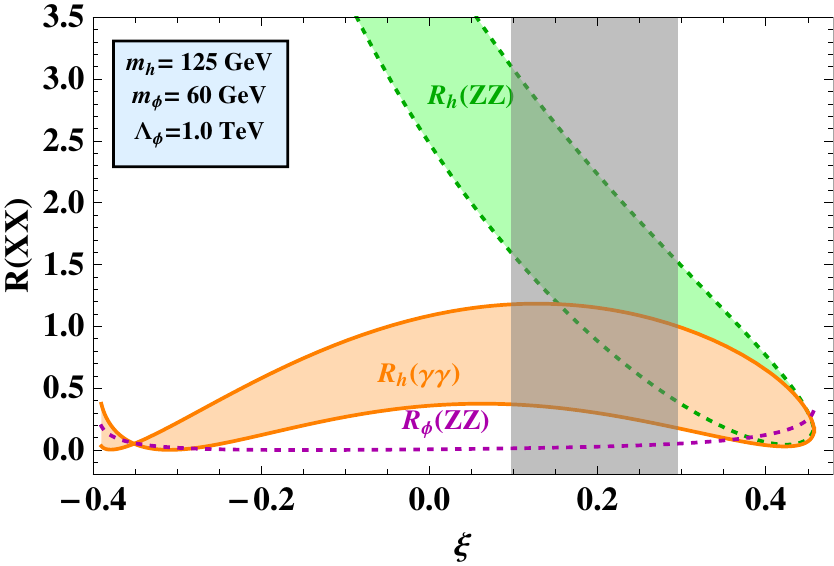}
&\hspace*{0.2cm}
	\includegraphics[width=2.6in,height=2.1in]{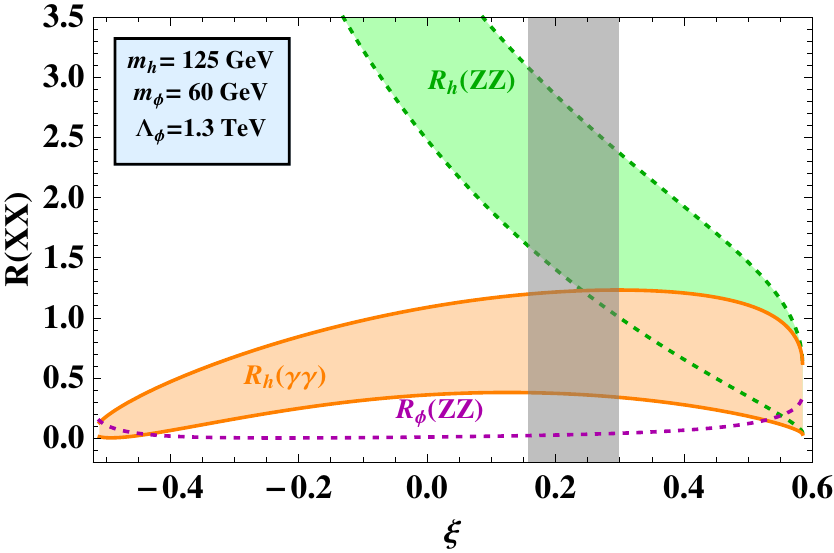}
	  \end{array}$ 
	   \hspace*{-0.2cm}
	\includegraphics[width=2.6in,height=2.1in]{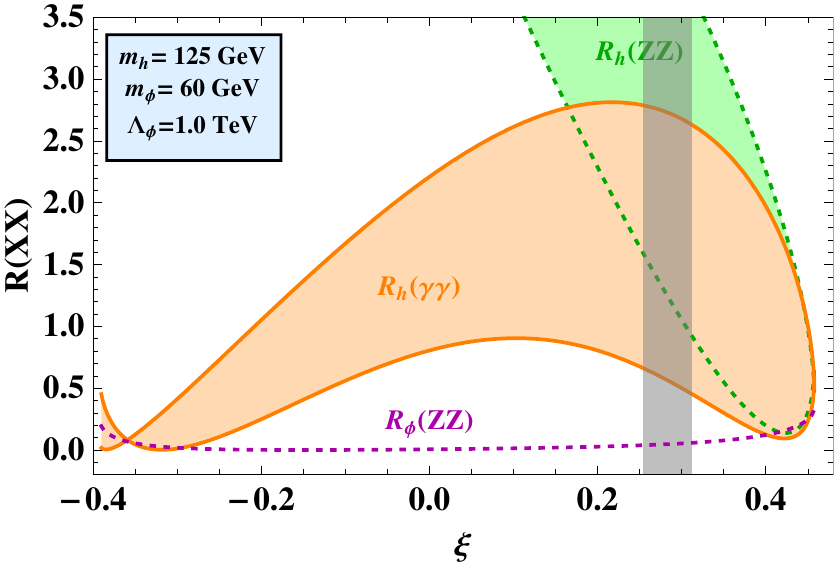}  
 \caption{Ratio of discovery significances $R(XX)\sim
   \sigma/\sigma_{SM}$, defined in the text, for $m_{h}=125$ GeV,
   $m_{\phi}=60$ GeV and for different values of $\Lambda_\phi$ and $m_{\tau'}$. For the upper
   panels we take  $m_{\tau'}=100$ GeV, while in the lower panel  we use $m_{\tau'}=150$ GeV. The light green bands indicate the
   theoretical uncertainties in the $gg\to h \to ZZ^*$ rate, while those for $\gamma \gamma$ are depicted in
   orange. The dashed purple lines marked by $R_\phi(ZZ)$ indicate the ratio
   of $\phi Z^*Z^*$ couplings with respect to the $h_{SM} Z^*Z^*$ one. The vertical gray bands
   indicate the allowed parameter space for $\xi$.}
\label{fig:mh125mphi60}
\end{figure}

\begin{itemize}

\item For Scenario 1a, if $m_\phi \le 100$ GeV, the LEP and Tevatron
  constraints apply. We find that, constraining $R_\phi(Z^*Z^*)$ to be in
  the required range ($<0.5$) forces $\xi<0.3$ and $R_h(ZZ^*)<1.6$. We
  varied the $\phi$ mass and found that for $m_\phi=60$ GeV  the
  experimental constraints (including LEP) are satisfied for
  $\Lambda_\phi =1.0$ TeV if $m_{\tau'}=150$ GeV, and for
  $\Lambda_\phi= 1.0, \,1.3$ TeV if $m_{\tau'}=100$ GeV. This is shown
  in Fig. \ref{fig:mh125mphi60}. The tight LEP constraints on the
  $\Lambda_\phi-\xi$ parameter space disallows greater
  values of $\Lambda_\phi$ in the very light $m_\phi$ parameter region. 

However, if $m_\phi=120$ GeV, there exist points in the parameter space
still consistent with all the experimental data for light $\tau'$ leptons. As both of the $h$ and
$\phi$ states are light, we graph the decays to $\gamma \gamma, ~b
{\bar b}$ and $ZZ^*$. This parameter point is a point in the CMS data, and may
or may not survive the latest round of data analysis.  As both Higgs-radion mixed
states are light, their branching ratios will depend on the $\tau'$
mass.  If $m_{\tau'}=100$ GeV, $\phi$ can decay into $\tau' \tau$, and
the branching ratios to $b{\bar b}, ZZ^*$ and $\gamma \gamma$ are modified. We
present these in Fig. \ref{fig:mh124mphi120} for $\Lambda_\phi=1.0,
1.3, \, 1.5$ and $2$ TeV.  From the figure, one can note that there
exist allowed regions of the parameter space for the last two
$\Lambda_\phi$ values only. Whereas for $m_{\tau'}=150$ GeV there are
no allowed bands in the parameter space which satisfy the constraints,  for
any $\Lambda_\phi$, as $R_h(ZZ^*)<1.6$. 
\begin{figure}[t]
\center$
	\begin{array}{cc}
	\includegraphics[width=2.6in,height=2.1in]{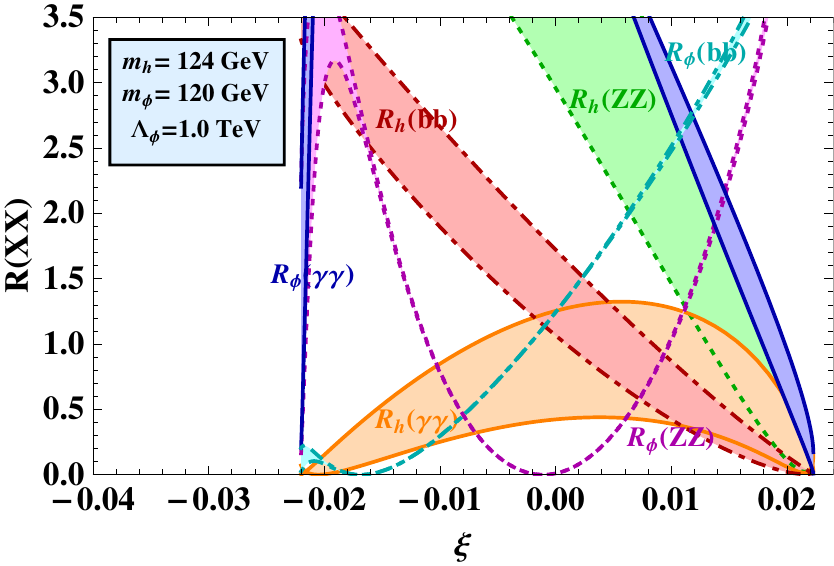}
&\hspace*{0.2cm}
	\includegraphics[width=2.6in,height=2.1in]{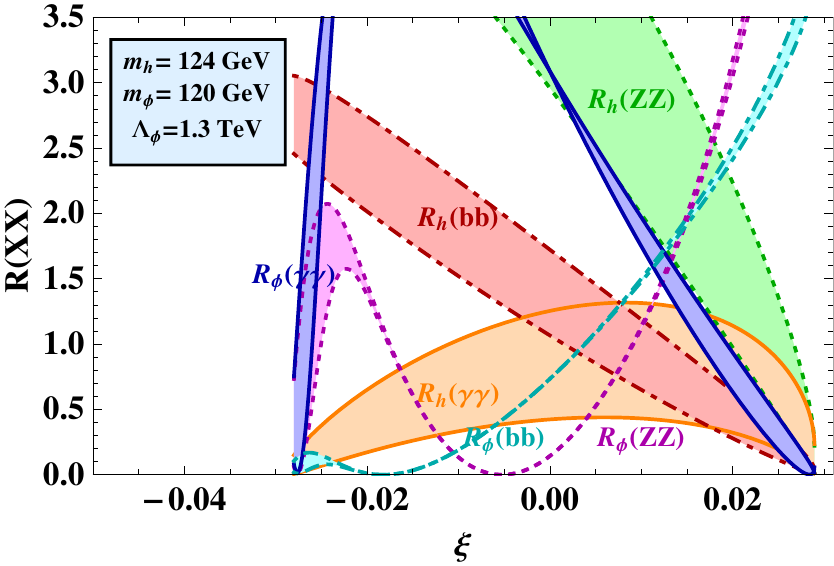}\\
	\includegraphics[width=2.6in,height=2.1in]{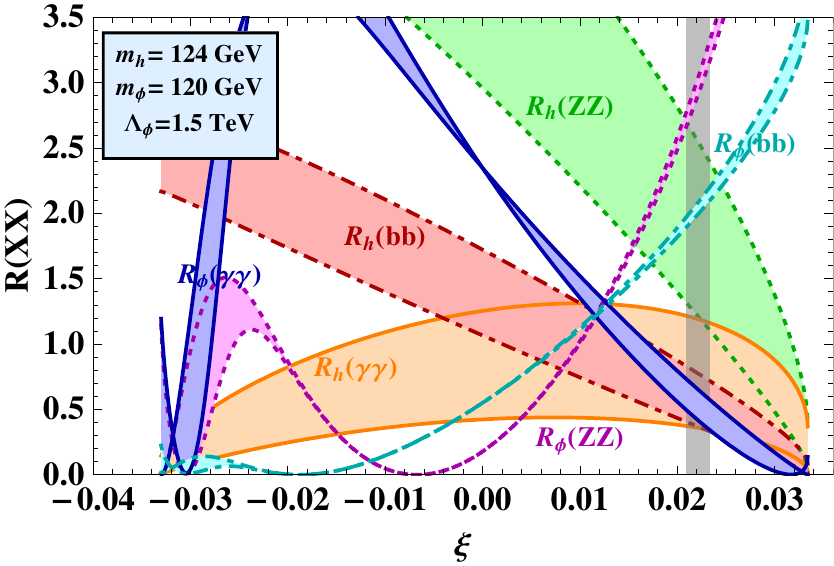}
&\hspace*{0.2cm}
	\includegraphics[width=2.6in,height=2.1in]{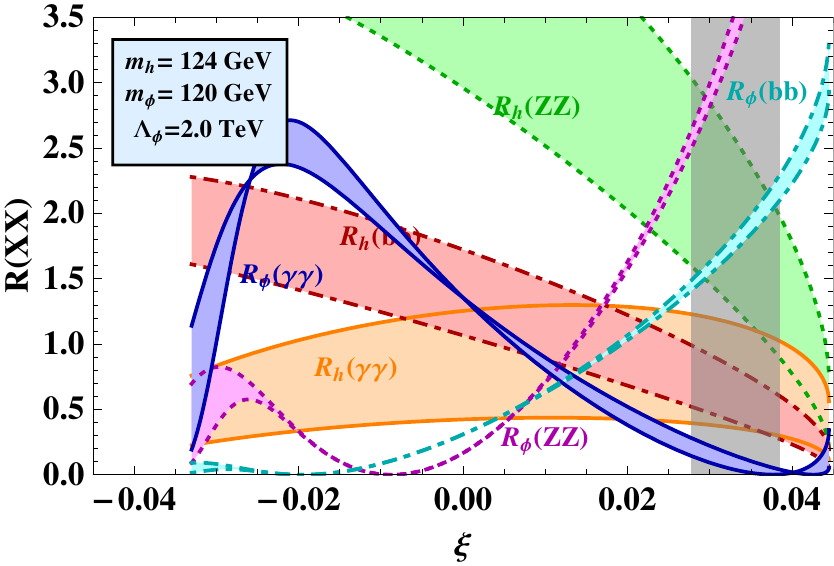}\\
        \end{array}$
 \caption{Ratio of discovery significances $R(XX)\sim
   \sigma/\sigma_{SM}$, defined in the text, for  $m_h=124$ GeV, $m_{\phi}=120$,
   $m_{\tau'}=100$ GeV and for different values of $\Lambda_\phi$. The light green
   bands indicate the theoretical uncertainties in the $ZZ^*$ signal,
   red for $b{\bar b}$ and orange for $\gamma \gamma$. For 
   $\phi$ the uncertainties are depicted in pink for $ZZ^*$, light blue
   for $b{\bar b}$ and purple for $\gamma \gamma$. The vertical gray bands
   indicate the allowed parameter space for $\xi$.}
\label{fig:mh124mphi120}
\end{figure}

If $m_h=124$ GeV, $m_\phi=137$ GeV, we are unable to find points in
the parameter space which satisfy the experimental constraints. If
$R_\phi(ZZ^*) <0.2$ as required, $R_\phi(\gamma \gamma)>2.3$, and
$R_h(ZZ^*)<1.6 $ for $\Lambda_\phi=1,\, 1.3,\, 1.5$ TeV, and the branching ratios worsen for higher $\Lambda_\phi$.  
\item For Scenario 1b, increasing $m_\phi$ only makes the situation
  worse and we do not find any region of parameters in which an $h$ state at  $125$
  GeV  and a heavy $\phi$ are allowed by the branching ratio
  constraints, and we thus choose not to show any figure for this case. 

We have so far found that only scenario 1a, allows some regions of
parameter space, but with a very restrictive $\Lambda_\phi$.

\item In Scenario 2a, where $m_{\phi}=125$ GeV and $h$ is light, and for $m_{\tau'}=150$ GeV, we do
  not find any allowed region in which all bounds and observed signals
  are respected. For regions where $R_h(\gamma \gamma)<0.5$, $R_\phi(ZZ^*)<1.6$. However, if the fourth generation
charged lepton $\tau'$ is light enough for the Higgs-radion mixed state(s) to
decay into it (through flavor-violating decays $\tau \tau'$), the
branching ratios are modified and the parameter space can shift. We
show this in Fig. \ref{fig:mtaup}, for $m_h=120$ GeV, $m_{\phi}=125$
GeV and $\Lambda_\phi=1$ TeV. The left-hand panel shows that even for $\Lambda_\phi=1.0$ TeV, there is no
allowed parameter space for $m_{\tau'}=150$ GeV. For $m_{\tau'}=100$ GeV, the possibility of
decays into $\tau \tau'$  reduces the branching ratios to the other
channels, thus widening the allowed $\xi$ parameter  range and of
$\Lambda_\phi$ for the Higgs-radion states. 
\begin{figure}[t]
\center$
	\begin{array}{cc}
	\includegraphics[width=2.6in,height=2.1in]{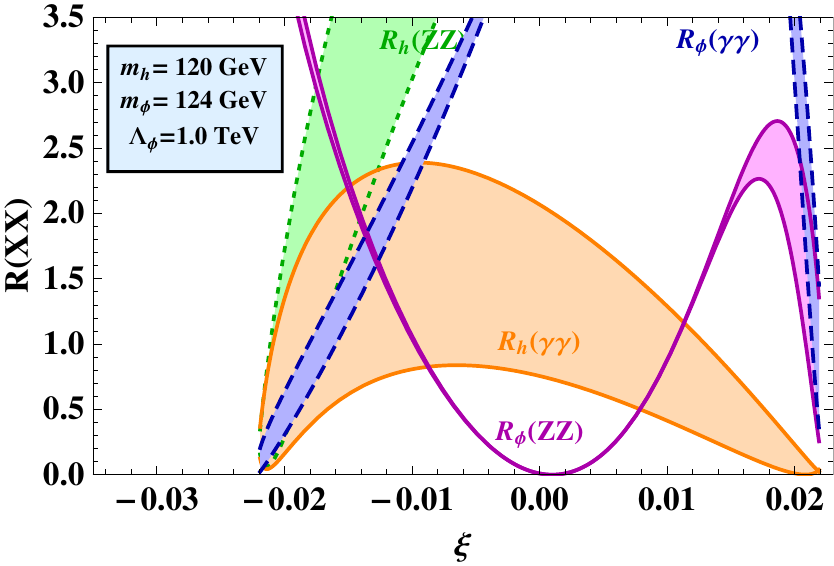}
&\hspace*{0.2cm}
	\includegraphics[width=2.6in,height=2.1in]{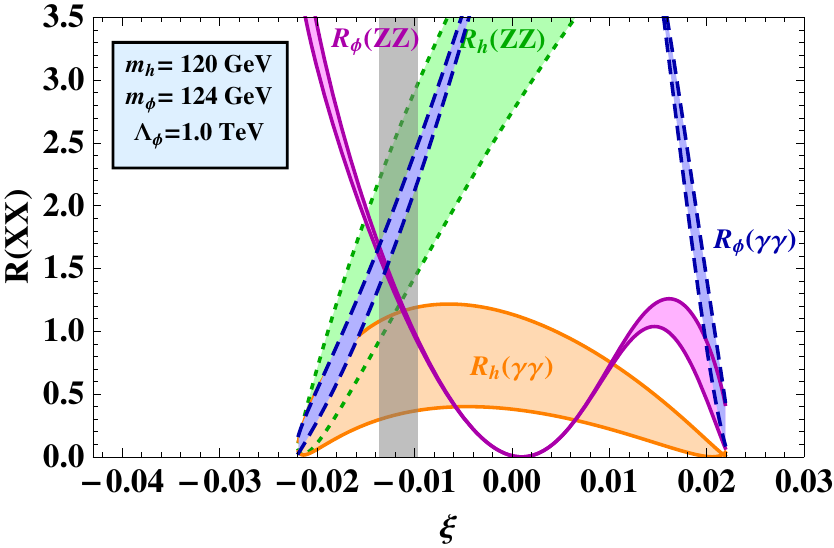}\\
	\includegraphics[width=2.6in,height=2.1in]{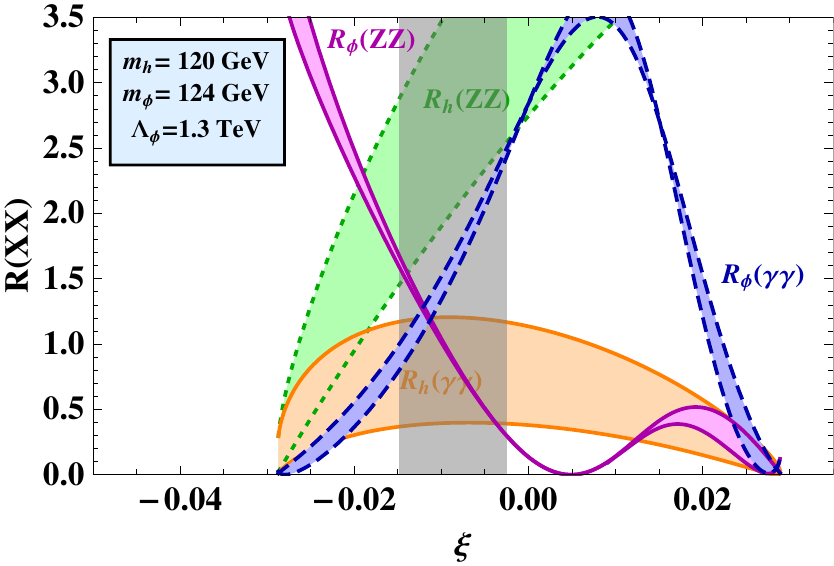}
&\hspace*{0.2cm}
	\includegraphics[width=2.6in,height=2.1in]{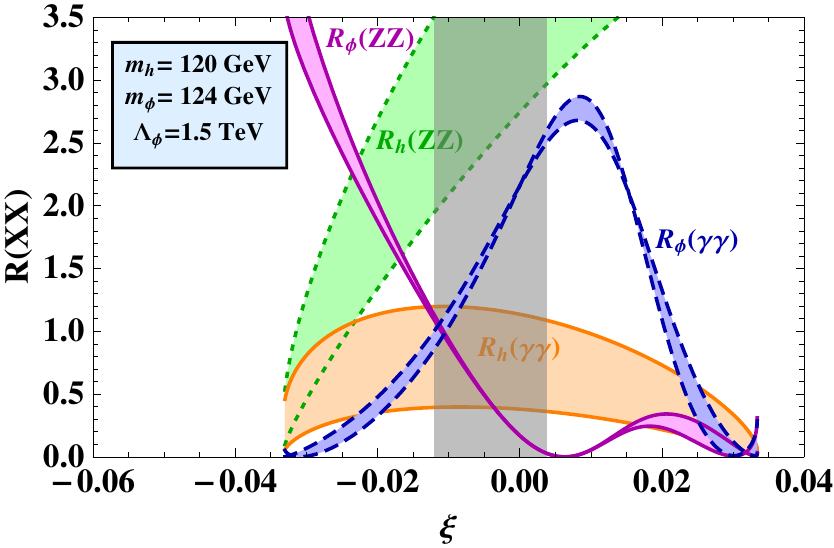}\\
	        \end{array}$
 \caption{Ratio of discovery significances $R(XX)\sim
   \sigma/\sigma_{SM}$, defined in the text, for $m_h=120$ GeV,
   $m_{\phi}=125$ GeV and for different
   values of $\Lambda_\phi$. In the top left-hand panel we took 
   $m_{\tau'}=150$ GeV and for the rest of panels we use $m_{\tau'}=100$ GeV. The light
   green bands indicate the theoretical uncertainties in 
the $ZZ^*$ signal and the orange the ones are for $\gamma
\gamma$.  For the $\phi$  the theoretical uncertainties in $ZZ^*$ are
given by pink bands and the ones for $\gamma \gamma$ are in
purple. The vertical gray bands indicate the allowed 
parameter space for $\xi$.}
\label{fig:mtaup}
\end{figure}

\item In Scenario 2b, on the other hand, we find that
  as long as $h$ is heavy enough, there are regions of parameter space
  where all experimental constraints are 
  satisfied. This is true independent of whether $m_{\tau'}=100$ or $150$ GeV. However, the results are quite sensitive to the value of
  $\Lambda_\phi$ and to the large experimental and theoretical
  uncertainties in the rates. We illustrate the situation for two values of
  $\Lambda_\phi$, i.e $\Lambda_\phi=1$ TeV in Fig. \ref{fig:mphi125L1000}
  and for $\Lambda_\phi=1.3$ TeV in Fig. \ref{fig:mphi125L1300}, for
  different $h$ masses  $m_h=320,\, 400,\, 500$ and $600$ GeV. Note
  that while for   $\Lambda_\phi=1$ TeV there are allowed bands for
  $600$, $500$ and    $320$ GeV, the parameter space for
  $\Lambda_\phi=1.3$ TeV is much more restrictive and we can only fit
  the data for $m_h=600$ GeV.   
\begin{figure}[t]
\center$
	\begin{array}{cc}
	\includegraphics[width=2.6in,height=2.1in]{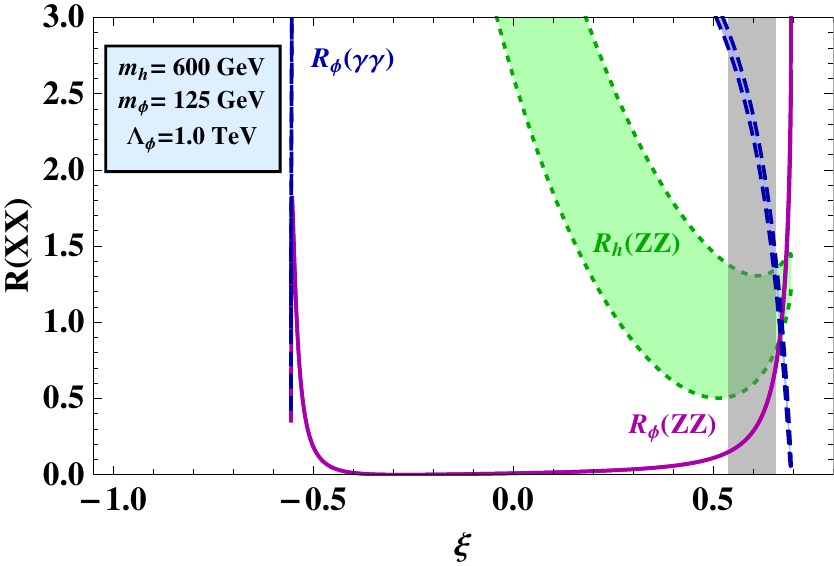}
&\hspace*{0.2cm}
	\includegraphics[width=2.6in,height=2.1in]{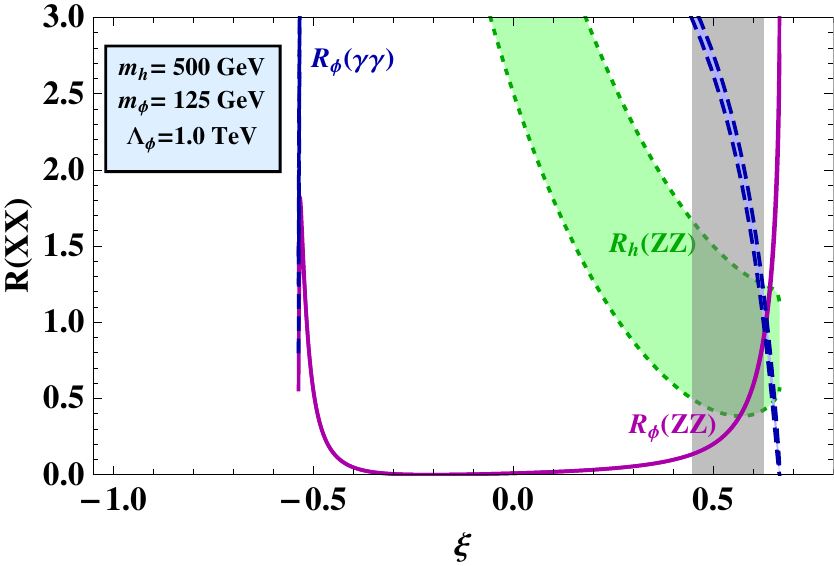}\\
	\includegraphics[width=2.6in,height=2.1in]{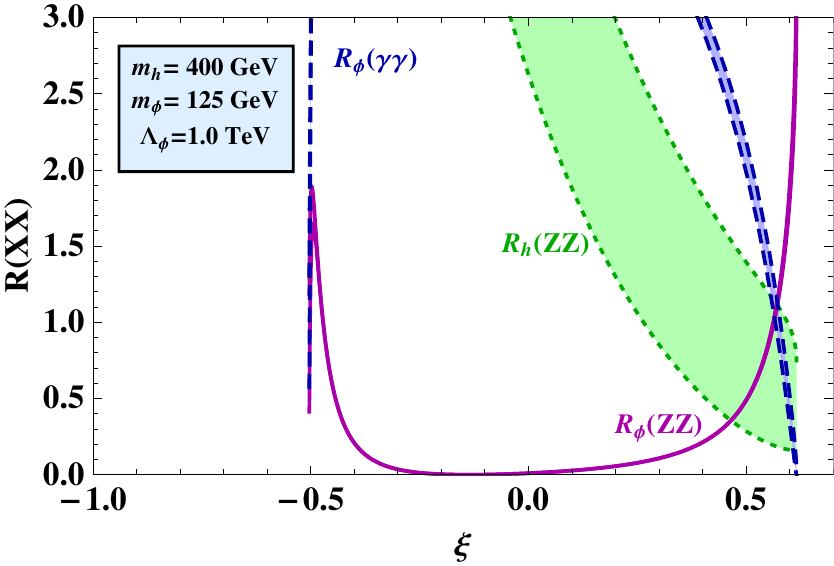}
&\hspace*{0.2cm}
	\includegraphics[width=2.6in,height=2.1in]{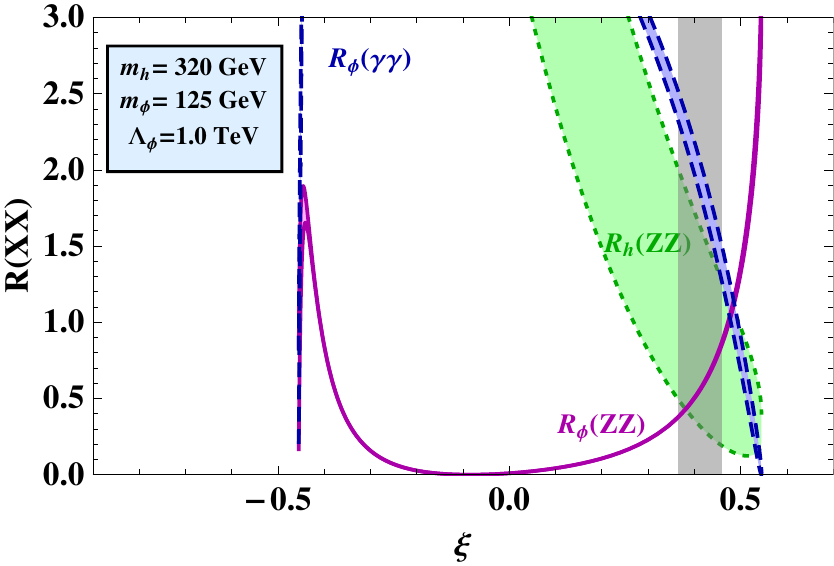}\\
        \end{array}$ 
\caption{Ratio of discovery significances $R(XX)\sim
   \sigma/\sigma_{SM}$, defined in the text, for $m_{\phi}=125$ GeV,
   $\Lambda_\phi=1.0$ TeV and for different masses of $h$. The light green bands indicate the theoretical uncertainties in
the $ZZ$ signal. We took $m_{\tau'}=150$ GeV, precluding FCNC decays to fourth generation leptons. The vertical gray bands indicate the allowed
parameter space for $\xi$.}
\label{fig:mphi125L1000}
\end{figure}
\begin{figure}[t]
\center$
	\begin{array}{cc}
	\includegraphics[width=2.6in,height=2.1in]{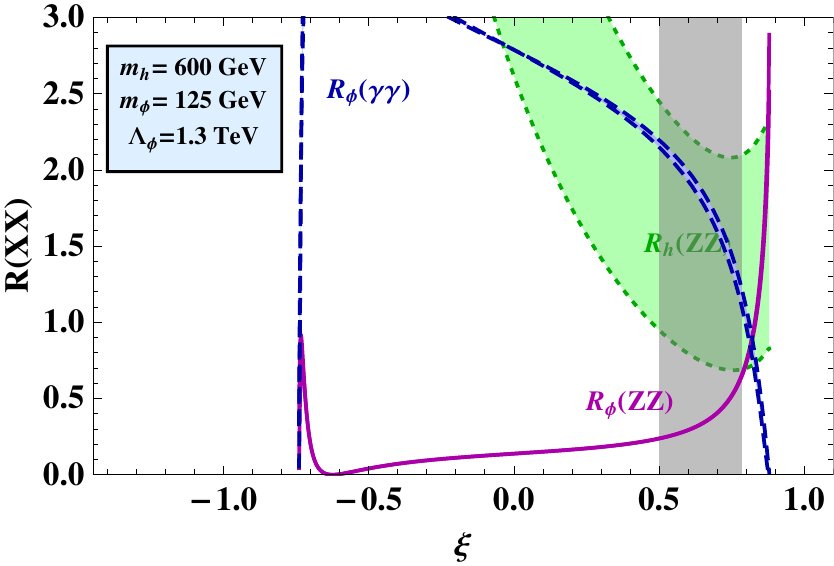}
&\hspace*{0.2cm}
	\includegraphics[width=2.6in,height=2.1in]{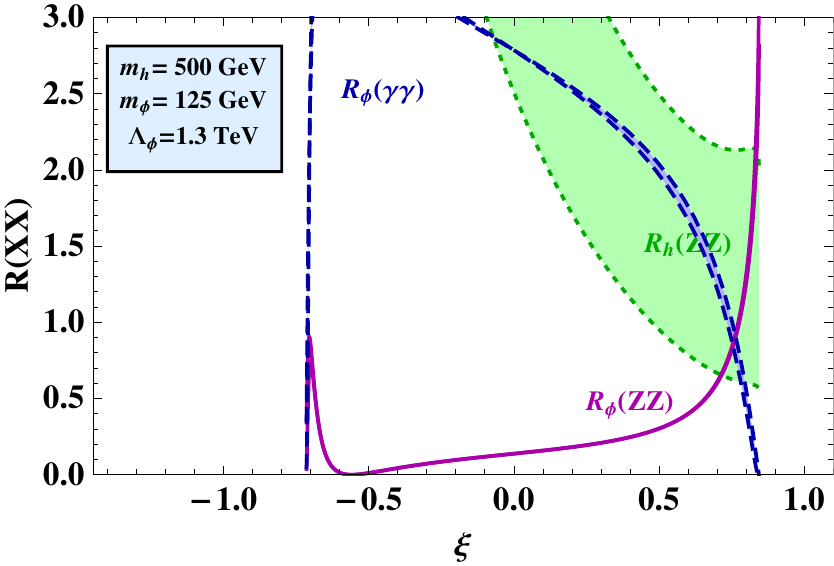}\\
	\includegraphics[width=2.6in,height=2.1in]{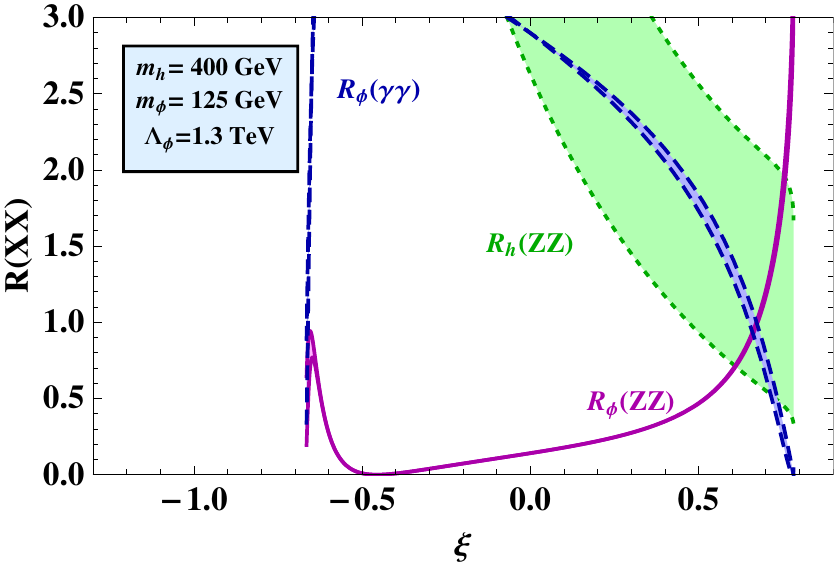}
&\hspace*{0.2cm}
	\includegraphics[width=2.6in,height=2.1in]{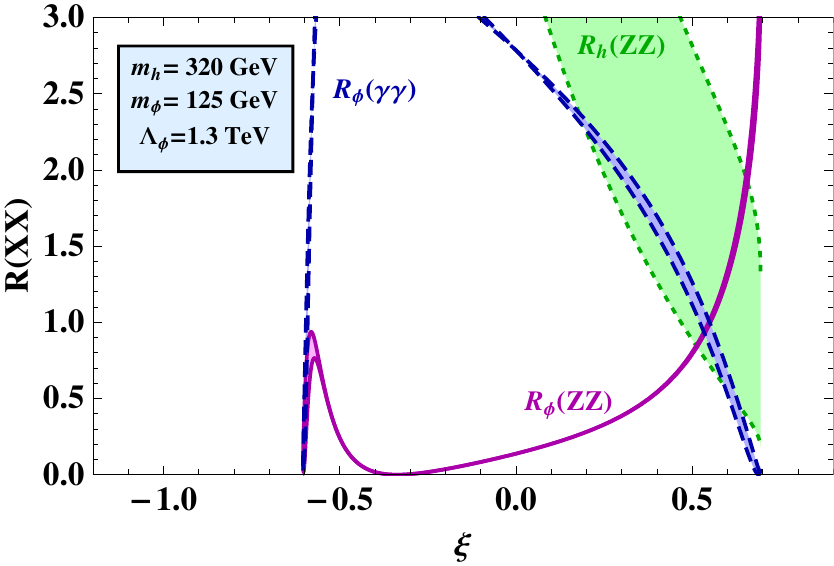}\\
        \end{array}$
 \caption{Same as Fig. \ref{fig:mphi125L1000}, but for  $\Lambda_\phi=1.3$ TeV.}
\label{fig:mphi125L1300}
\end{figure}
\end{itemize}
%

\section{Flavor Changing Decays of the Higgs-radion states in the four Generation Model}
\label{sec:FCNC}

Should the scalar discovered at the LHC be a Higgs-radion mixed state, its decay into two fermions will be different than for a SM Higgs boson, and further analysis at the LHC could differentiate the particles. In this section we  present the branching ratios of the mixed
Higgs-radion state into two fermions. We start by giving
analytical formulas, as they have not appeared before, then show
specific values for the flavor-conserving and the flavor-violating
branching ratios, for the allowed  points in the parameter space
presented in the previous section. The branching ratios of the mixed
states $\phi$ and $h$ into two fermions are given by:
\begin{eqnarray}
 \Gamma(\phi\rightarrow \bar{f}_if_j)&=&\frac{Sc}{16\pi
m_{\phi}^3}\sqrt{m_{\phi}^4+m_i^4+m_j^4-2m_{\phi}^2m_i^2-2m_{\phi}^2m_j^2
-2m_i^2m_j^2} \nonumber\\
&\times&\frac{2 m_i
m_j}{\upsilon^2}\Bigg(\Big[({\tilde c}_{ij}
)^2+({\tilde c}_{ji})^2\Big](-m_{\phi}
^2+m_i^2+m_j^2)
+4~\Re\big[({\tilde c}_{ij})({\tilde c}_{ji})\big]m_i m_j\Bigg),
\end{eqnarray}
\begin{eqnarray}
 \Gamma(h\rightarrow \Bar{f}_if_j)&=&\frac{Sc}{16\pi
m_{h}^3}\sqrt{m_{h}^4+m_i^4+m_j^4-2m_{h}^2m_i^2-2m_{h}^2m_j^2
-2m_i^2m_j^2} \nonumber\\
&\times&\frac{2 m_i
m_j}{\upsilon^2}\Bigg(\Big[({\tilde d}_{ij}
)^2+({\tilde d}_{ji})^2\Big](-m_{h}
^2+m_i^2+m_j^2)
+4~\Re\big[({\tilde d}_{ij})({\tilde d}_{ji})\big]m_i m_j\Bigg).
\end{eqnarray}
Here $S$ is a product of statistical factors $1/j!$ for each group of
$j$ identical particles in the final state. For flavor violating couplings, the
particles in the final state are different, therefore, $S=1$. The factor $c$ is
the color factor, for quarks $c=3$, and for leptonic decays,
$c=1$. 

The flavor violating couplings of the mixed states are defined as 
\begin{eqnarray}
{\tilde c}_{ij}&=&c\ a_{ij}+a\gamma\ \tilde{a}_{ij}, \nonumber \\
{\tilde d}_{ij}&=&d\ a_{ij}+b\gamma\ \tilde{a}_{ij},
\end{eqnarray}
where the couplings $a_{ij}$ and $\tilde{a}_{ij}$, of the original
unmixed Higgs and radion, have been previously obtained in
\cite{Azatov:2009na,Azatov:2008vm} in the case of three generations
and in \cite{Frank:2011rw,Frank:2011kz} with four generations.
In the branching ratio calculations given in the Tables, we use the central values for $a_{ij}$s and
$\tilde{a}_{ij}$s obtained in the numerical scans performed in the
last references, and we choose a specific allowed value of $\xi$ for
each point studied in the parameter space. 

We first present the branching ratios to FCNC decays for allowed parameter points from the previous section. We chose two different scenarios. In one $m_{\tau'}=100$ GeV, thus a scalar of mass $125$ GeV can have flavor-violating decays into $\tau \tau'$. These results are shown in Table I.
The FCNC decay branching ratios into $\tau \tau'$ can reach 5\%. Overall, the effect is not measurable, however, should the mass of the $\tau'$ be close to its experimental limit $100$ GeV, the situation could change drastically and the $BR (\phi \to \tau' \tau)$ can reach 50 \%, suppressing all other decays.

\begin{table}
\footnotesize
\begin{tabular}{ |c | c | c | c | c | c | c | c | c | }
\hline
  $\mathbf{\Lambda (TeV)}$ & $\boldsymbol{\xi}$ & $\mathbf{m (GeV)}$ &
$\mathbf{b'b}$ & $\mathbf{tc}$ & $\mathbf{bs}$ &
  $\boldsymbol{\tau'\tau}$ & $\boldsymbol{\mu\tau}$ &
$\boldsymbol{\nu_{\tau}\nu_{\tau'}}$\\
\hline\hline
 \multirow{2}{*}{1.3} &  \multirow{2}{*}{0.228} & $m_{\phi}=60$& - & - &
$9.53\times 10^{-5}$ & - & $2.49\times 10^{-5}$ & -\\
  &  & $m_{h}=125$ & - & - & $5.51\times 10^{-4}$ & $5.53\times 10^{-1}$ &
$1.52\times 10^{-4}$ & $1.46\times 10^{-3}$\\
\hline\hline
 \multirow{2}{*}{1.3} &  \multirow{2}{*}{-0.00866} & $m_{\phi}=124$& - & - &
$7.30\times 10^{-5}$ & $4.67\times 10^{-2}$ & $2.01\times 10^{-5}$ &
$7.76\times 10^{-4}$\\
  &  & $m_{h}=120$ & - & - & $6.34\times 10^{-4}$ & $4.74\times 10^{-1}$ &
$1.77\times 10^{-4}$ & $1.48\times 10^{-3}$\\
\hline\hline
 \multirow{2}{*}{1.5} &  \multirow{2}{*}{0.0221} & $m_{\phi}=120$& - & - &
$3.31\times 10^{-4}$ & $2.20\times 10^{-1}$ & $9.06\times 10^{-5}$ &
$1.46\times 10^{-3}$\\
  &  & $m_{h}=124$ & - & - & $7.34\times
10^{-4}$ & $7.19\times 10^{-1}$ & $2.02\times 10^{-4}$ & $1.76\times 10^{-3}$\\
\hline\hline
  \multirow{2}{*}{1.0} & \multirow{2}{*}{0.417} & $m_{\phi}=125$ & - & - &
$7.47\times 10^{-5}$ & $5.94\times 10^{-2}$ & $2.05\times 10^{-5}$ &
$8.15\times 10^{-4}$\\
  &  & $m_{h}=320$ & - & $1.53\times 10^{-4}$ & $1.04\times
10^{-6}$ & $7.84\times 10^{-3}$ & $3.23\times 10^{-7}$ & $9.28\times 10^{-6}$\\
\hline\hline
  \multirow{2}{*}{1.0} & \multirow{2}{*}{0.537} & $m_{\phi}=125$ & - & - &
  $4.21\times 10^{-5}$ & $2.95\times 10^{-3}$ & $1.16\times 10^{-5}$ &
$6.92\times 10^{-4}$\\
  &  & $m_{h}=500$ & $1.27\times 10^{-2}$ &$6.61\times 10^{-5}$ & $2.93\times
10^{-7}$ & $2.71\times 10^{-3}$ & $9.83\times 10^{-8}$ & $3.00\times 10^{-6}$\\
\hline\hline
  \multirow{2}{*}{1.0} & \multirow{2}{*}{0.601} & $m_{\phi}=125$ & - & - &
  $1.52\times 10^{-5}$ & $7.29\times 10^{-3}$ & $4.17\times 10^{-6}$ &
$5.42\times 10^{-4}$\\
  &  & $m_{h}=600$ & $1.44\times 10^{-2}$ &$4.84\times 10^{-5}$ & $1.99\times
10^{-7}$ & $1.89\times 10^{-3}$ & $6.68\times 10^{-8}$ & $2.02\times 10^{-6}$\\
\hline
\end{tabular}
\label{tbl:FCNC}
\caption{The FCNC branching ratios of $h$ and $\phi$ for allowed points in the parameter space.
The fourth generation fermion masses are chosen as $m_{t'}=400$ GeV,
$m_{b'}=350$ GeV, $m_{\tau'}=100$ GeV, $m_{\nu_{\tau'}}=90$ GeV. }
\end{table}

In Table II,
 we chose $m_{\tau'}=150$ GeV, precluding FCNC decays of the lightest scalar into fourth generation leptons. As before, the Higgs-radion mixed state can decay into third and fourth generation neutrinos, but the branching ratios are not significant. For the other fourth generation fermions, we take throughout  $m_{t'}=400$ GeV,
$m_{b'}=350$ GeV, and $m_{\nu_{\tau'}}=90$ GeV.

\begin{table}
\footnotesize
\begin{tabular}{ |c | c | c | c | c | c | c | c | c | }
\hline
  $\mathbf{\Lambda (TeV)}$ & $\boldsymbol{\xi}$ & $\mathbf{m (GeV)}$ &
$\mathbf{b'b}$ & $\mathbf{tc}$ & $\mathbf{bs}$ &
  $\boldsymbol{\tau'\tau}$ & $\boldsymbol{\mu\tau}$ &
$\boldsymbol{\nu_{\tau}\nu_{\tau'}}$\\
\hline\hline
 \multirow{2}{*}{1.0} &  \multirow{2}{*}{0.0283} & $m_{\phi}=60$& - & - &
$1.08\times 10^{-5}$ & - & $2.83\times 10^{-6}$ &-\\
  &  & $m_{h}=125$ & - & - & $1.05\times
10^{-3}$ & - & $2.88\times 10^{-4}$ & $3.00\times 10^{-3}$\\
\hline\hline
  \multirow{2}{*}{1.0} & \multirow{2}{*}{0.412} & $m_{\phi}=125$ & - & - &
$7.60\times 10^{-5}$ & - & $2.09\times 10^{-5}$ &
$8.52\times 10^{-4}$\\
  &  & $m_{h}=320$ & - & $1.61\times 10^{-4}$ & $1.10\times
10^{-6}$ & $9.24\times 10^{-3}$ & $3.40\times 10^{-7}$ & $9.76\times 10^{-6}$\\
\hline\hline
  \multirow{2}{*}{1.0} & \multirow{2}{*}{0.565} & $m_{\phi}=125$ & - & - &
  $5.84\times 10^{-5}$ & - & $1.61\times 10^{-5}$ &
$7.85\times 10^{-4}$\\
  &  & $m_{h}=500$ & $1.26\times 10^{-2}$ &$6.51\times 10^{-5}$ & $2.90\times
10^{-7}$ & $3.62\times 10^{-3}$ & $9.72\times 10^{-8}$ & $2.91\times 10^{-6}$\\
\hline\hline
  \multirow{2}{*}{1.0} & \multirow{2}{*}{0.644} & $m_{\phi}=125$ & - & - &
  $4.50\times 10^{-5}$ & - & $1.24\times 10^{-5}$ &
$7.28\times 10^{-4}$\\
  &  & $m_{h}=600$ & $1.42\times 10^{-2}$ &$4.76\times 10^{-5}$ & $1.96\times
10^{-7}$ & $2.59\times 10^{-3}$ & $6.57\times 10^{-8}$ & $1.99\times 10^{-6}$\\
\hline
\end{tabular}
\label{tbl:FCNC2}
\caption{Same as Table I, but for  $m_{\tau'}=150$ GeV. }
\end{table}
%
We perform the same analysis, this time for the flavor-diagonal couplings, in Table III
for $m_{\tau'}=100$ GeV and in Table IV
for $m_{\tau'}=150$ GeV. As no flavor-conserving decays into fourth
generation fermions are possible, we compare the ratio of significance and Yukawa couplings to
the corresponding ones in the SM. The light scalar state (at $120$ or
$125$ GeV) exhibits large enhancements for $b{\bar b}$
and $c{\bar c}${\footnote{The enhancements in $b {\bar b}$ for the $\phi$ state are consistent with the latest Tevatron results $R(b{\bar b})=2.03^{+0.73}_{-0.71}$ \cite{TevatronMoriond}.}}, while the heavier scalars have correspondingly suppressed
ratios of significance with respect to the SM. The enhancements are
inherited from the couplings of the Higgs boson \cite{Frank:2011rw},
and if observed, they will give a clear indication for the warped
space model.
 %
\begin{table}
\footnotesize
\begin{tabular}{ |c | c | c | c | c | c | c |}
\hline
  $\mathbf{\Lambda (TeV)}$ & $\boldsymbol{\xi}$ & $\mathbf{m (GeV)}$ &
  $\mathbf{R(bb)}$ & $\mathbf{R(cc)}$ & $\mathbf{R(tt)}$ & $\mathbf{Y_{tt}}$ \\
\hline\hline
\multirow{2}{*}{1.5} &  \multirow{2}{*}{0.0221} & $m_{\phi}=120$ & 2.05 & 2.13
& -&0.496\\
  &  & $m_{h}=124$ & 0.563 & 0.557 & -& 0.880\\
\hline\hline
\multirow{2}{*}{1.0} &  \multirow{2}{*}{0.421} & $m_{\phi}=125$ & 2.20 & 2.31
& -&0.380\\
  &  & $m_{h}=320$ & 0.523 & 0.513 & -& 1.02\\
\hline\hline
\multirow{2}{*}{1.0} &  \multirow{2}{*}{0.537} & $m_{\phi}=125$ & 1.81 & 1.93
& -& 0.317\\
  &  & $m_{h}=500$ & 0.532 & 0.521 & 0.556 & 1.19\\
\hline\hline
\multirow{2}{*}{1.0} &  \multirow{2}{*}{0.601} & $m_{\phi}=125$ & 1.78 &
1.89 & -& 0.316\\
  &  & $m_{h}=600$ & 0.553 & 0.520 & 0.559 & 1.36\\
\hline
\end{tabular}
\label{tbl:Yukawa}
\caption{Ratio of significance $R_{h(\phi)}(XX)=S(gg\rightarrow
h(\phi)\rightarrow f\bar{f})\slash S(gg\rightarrow h_{SM}\rightarrow
f\bar{f})$ for different
parameter space. Last column are the Yukawa couplings for $h(\phi)$ to
$t\bar{t}$. The fourth generation fermion masses are chosen as $m_{t'}=400$ GeV,
$m_{b'}=350$ GeV, $m_{\tau'}=100$ GeV, $m_{\nu_{\tau'}}=90$ GeV.}
\end{table}

%
\begin{table}
\footnotesize
\begin{tabular}{ |c | c | c | c | c | c | c |}
\hline
  $\mathbf{\Lambda (TeV)}$ & $\boldsymbol{\xi}$ & $\mathbf{m (GeV)}$ &
  $\mathbf{R(bb)}$ & $\mathbf{R(cc)}$ & $\mathbf{R(tt)}$ & $\mathbf{Y_{tt}}$ \\
\hline\hline
\multirow{2}{*}{1.0} &  \multirow{2}{*}{0.412} & $m_{\phi}=125$ & 2.45 & 2.59
& -& 0.374\\
  &  & $m_{h}=320$ & 0.586 & 0.575 & -& 1.01\\
\hline\hline
\multirow{2}{*}{1.0} &  \multirow{2}{*}{0.565} & $m_{\phi}=125$ & 2.02 & 2.14
& -& 0.334\\
  &  & $m_{h}=500$ & 0.481 & 0.470 & 0.503 & 1.24\\
\hline\hline
\multirow{2}{*}{1.0} &  \multirow{2}{*}{0.480} & $m_{\phi}=125$ & 1.48 &
1.59 & -& 0.602\\
  &  & $m_{h}=600$ & 0.636 & 0.625 & 0.661 & 0.819\\
\hline
\end{tabular}
\label{tbl:Yukawa2}
\caption{Same as Table III, but for  $m_{\tau'}=150$ GeV.}
\end{table}


\section{Conclusions and Outlook}
\label{sec:conclusion}

In this work, we have investigated the phenomenology of the
Higgs-radion mixed state with a fourth generation of quarks and
leptons, in an attempt to explain the latest LHC data. We asked the
question: if the scalar particle seen at the LHC is not the ordinary
SM Higgs boson, but a mixed Higgs-radion state, could this state
satisfy all the experimental constraints, even including the effects
of a fourth generation? The four generation assumption in warped space
models is of particular interest, as  the Standard Model with four generations, SM4, fails to reproduce the
observed data to at least 95\% confidence level. 
A fourth generation, which is severely restricted and perhaps even ruled
out by the ATLAS and CMS data in SM4 could be resuscitated in warped space
models. The answer to the question we posed is a cautious yes. That
is, there exist regions of the parameter space where one of the mixed
Higgs-radion states has mass of $125$ GeV, and satisfies existing
experimental constraints, while the other either has a mass of $120$
GeV, thus fitting 
a CMS parameter point, or evades present
collider bounds.  

Summarizing the allowed parameter space, if the $h$ state is the scalar
observed at the LHC, the $\phi$ mass must be light. Parameter points
with either $m_\phi=60$ GeV, which evade LEP restrictions, or
$m_\phi=120$ GeV, which fit the CMS data, are allowed for some range of
the mixing parameter $\xi$. We analyzed these for both very light
fourth generation charged leptons, $m_{\tau'}=100$ GeV, or for heavier
ones, $m_{\tau'}=150$ GeV. The difference between these two masses is that
the first case allows flavor-changing decays of the Higgs-radion state,
which are large in this model and which modify the branching ratios to
$\gamma \gamma$ and $ZZ^*$. All of these parameter points require the
scale $\Lambda_\phi$ to be light, in the $1.0 -1.3$ TeV range, the
exact values dependent on the rest of the parameters. For larger
$m_\phi$ values, the branching ratio to $ZZ$ increases beyond the LHC
limits, and thus this parameter region is forbidden.  This region of
parameter space is very fragile. For $m_h=124$ GeV, the point at $
m_\phi=120$ GeV shows signs of instability as the $4\ell$ excess might
be cancelled by $\gamma \gamma$, while its decay into $b{\bar b}$ appears to
have increased.
The signal for $m_\phi=60$ GeV, while 
not ruled out by LEP data depends very sensitively on the values of $m_{\tau'}$ and $\Lambda_\phi$. 

If $\phi$ is the scalar observed at the LHC, the $h$ state is most
likely to be heavy. The exception is when  $m_{\tau'}=100$ GeV; for
$m_h=120$ GeV  parameter points exist for $\Lambda_\phi=1.0,\,1.3,$
and $1.5$ TeV. Regions where $m_h=320, \,400,\, 500$ and $600$ 
GeV exist for some values of $\Lambda_\phi$, which is still required
to be in the $1.0-1.5$ TeV range.  These parameter regions seem quite
robust and not dependent on whether $\tau'$ is heavy or light; however
they could be ruled out within the next year at LHC as data for
heavier scalars becomes available.  
To increase predictability of our scenario, we calculated the branching ratios of the
allowed Higgs-radion states into fermions, both for flavor changing
and flavor conserving channels (some of which are significantly
enhanced with respect to the SM expectations). As more data on the
scalar production and decay becomes available, these predictions can
be compared with the experiment, specially noting the appearance of
the interesting exotic FCNC decays.  

In conclusion, we have achieved two goals in this work: first, we have
shown that a scalar in a warped model with a fourth generation of fermions can be light {\it and}
consistent with the  LHC data, if the observed particle is a
Higgs-radion mixed state. 
Second, the allowed parameter space is tightly constrained and
expected to be confirmed or ruled out within a year by  further
analyses and/or higher luminosity at the LHC.

\section{Acknowledgments}

 We acknowledge NSERC of Canada for partial financial
support under grant number SAP105354.



\end{document}